\documentclass[prl,reprint,superscriptaddress]{revtex4-1}

\usepackage{latexsym,amssymb,amstext,amsmath}
\usepackage{slashed}
\usepackage{mathrsfs}
\usepackage{color}
\usepackage{hyperref}
\usepackage{verbatim}
\usepackage{graphicx}
\usepackage{color}
\usepackage{mathrsfs}

\newcommand {\cL}{{\cal L}}

\def\a{\alpha}
\def\b{\beta}

\def\d{\delta}
\def\e{\epsilon}

\def\g{\gamma}

\def\l{\lambda}
\def\m{\mu}
\def\n{\nu}
\def\o{\omega}

\def\r{\rho}
\def\s{\sigma}
\def\t{\tau}

\def\O{\Omega}

\newcommand{\ve}{\varepsilon}

\newcommand{\be}{\begin{equation}}
	\newcommand{\ee}{\end{equation}}
\newcommand{\bea}{\begin{eqnarray}}
	\newcommand{\eea}{\end{eqnarray}}

\newcommand{\ba}{\begin{array}}
	\newcommand{\ea}{\end{array}}

\def\double #1{#1{\hbox{\kern-2pt $#1$}}}

\newcommand{\bsubeq}{\begin{subequations}}
	\newcommand{\esubeq}{\end{subequations}}

\begin{document}

\title{Three-Dimensional Extended Newtonian (Super)Gravity}

\author{Nese Ozdemir}
\email{nozdemir@itu.edu.tr}
\affiliation{Department of Physics,
	Istanbul Technical University,
	Maslak 34469 Istanbul,
	Turkey}

\author{Mehmet Ozkan}
\email{ozkanmehm@itu.edu.tr}
\affiliation{Department of Physics,
Istanbul Technical University,
Maslak 34469 Istanbul,
Turkey}

\author{Orhan Tunca}
\email{tuncaor@itu.edu.tr}
\affiliation{Department of Physics,
	Istanbul Technical University,
	Maslak 34469 Istanbul,
	Turkey}

\author{Utku Zorba}
\email{zorba@itu.edu.tr}
\affiliation{Department of Physics,
	Istanbul Technical University,
	Maslak 34469 Istanbul,
	Turkey}

	

\begin{abstract}
We present a three dimensional non-relativistic model of gravity that is invariant under the central extension of the symmetry group that leaves the recently constructed Newtonian gravity action invariant. We show that the model arises from the contraction of a bi-metric model that is the sum of the Einstein gravity in Lorentzian and the Euclidean signatures. We also present the supersymmetric completion of this action which provides one of the very few examples of an action for non-relativistic supergravity.

\end{abstract}
	

\maketitle
\allowdisplaybreaks

Newton-Cartan geometry has been studied extensively in recent years due to its use in condensed matter systems and non-AdS holography \cite{CM1,CM2,CM3,CM4,HL1,HL2,HL3,HL4,HL5,HL6,HL7}. This geometric setup is the necessary framework to covariantize the Poisson equation of Newtonian gravity with the underlying symmetries are given by the centrally extended Galilei algebra, known as the Bargmann algebra \cite{Cartan1,Cartan2,Trautman}. While the Poisson equation was geometrized using Newton-Cartan geometry long ago, an action principle for the Newtonian gravity has only recently been constructed \cite{NewtonAction}. The major feature of this action is that although the Bargmann algebra can capture the symmetries of the required field equation, it is not capable to obtain the Newtonian gravity action but an extension of that algebra with additional three new generators is necessary. It was, furthermore, mentioned in \cite{NewtonAction} that this new algebra can be obtained by a contraction of the direct sum of the Poincar\'e and Euclidean algebras in $D$ dimensions.

In this paper, we shall focus on the properties of the symmetry algebra that gives rise to Newtonian gravity action in three dimensions. The reason for that is twofold. First, three-dimensional non-relativistic geometries have found themselves direct applications in condensed matter physics \cite{CM2, CM3,3dCM1,3dCM2,3dCM3}. Therefore, such new symmetries would presumably be expected to be relevant to a realistic theory. Second, gravity, both relativistically and non-relativistically, is technically much simpler to handle in three dimensions as it can be treated as a Chern-Simons gauge theory \cite{3d1,3d2,EBG,EMG0,EBG2,MCSG}. This technical advantage particularly allows us to immediately establish the three-dimensional action and to overcome two major difficulties with non-relativistic model construction, namely to obtain the model by taking a non-relativistic limit and to find the supersymmetric completion. In particular, we show that the three-dimensional model is the contraction of a bi-metric model that is the sum of the Einstein gravity in Lorentzian and the Euclidean signatures. Here, it is also worthwhile to briefly discuss the current status of non-relativistic supergravity. While supergravity in its relativistic formulation is now a well-established subject, its non-relativistic formulation is far from being complete with only very recent constructions \cite{EBG,SGR1,SGR2}. Such models are particularly relevant to the construction of supersymmetric field theories on curved backgrounds by means of localization \cite{Loc1,Loc2}. Here, we provide the supersymmetric completion of the extended Newtonian gravity, which allows for such supersymmetric curved backgrounds, similar to extended Bargmann supergravity \cite{EBG}.

The spacetime symmetry algebra that is needed to construct an action principle for Newtonian gravity consists of the generators of the Bargmann algebra $\{H, P_a, G_a, J_{ab}, M\}$ as well as a set of additional generators $\{T_{a}, B_a, S_{ab}\}$ \cite{NewtonAction}. To construct a Chern-Simons action based on this algebra in three-dimensions, one needs to introduce a central extension in order to have a non-degenerate bilinear form. Here, we introduce two central charges, $Y$ and $Z$, and the non-zero commutation relations for the extended algebra are given by
\bea
&& \left[ H, G_a \right]  = - \e_{ab} P^b \,,  \left[ J, P_a \right] = - \e_{ab} P^b \,, \left[ J, G_a \right] =- \e_{ab} G^b \,,\nonumber\\
&& \left[ G_a, P_b \right] =  \e_{ab}M \,, \left[ G_a, G_b \right] =  \e_{ab} S \,,  \quad \left[ J, B_a \right] = - \e_{ab} B^b \,, \nonumber\\
&& \left[ J, T_a \right] =- \e_{ab} T^b \,,  \left[ H, B_a \right] = - \e_{ab} T^b \,,  \left[ M, G_a \right] = - \e_{ab} T^b \,,\nonumber\\
&& \left[ S, P_a \right] =  -\e_{ab} T^b \,,  \left[ S, G_a \right] =  -\e_{ab} B^b \,,  \left[ G_a, T_b \right] = - \e_{ab}Y \,,\nonumber\\
&& \left[ G_a, B_b \right] =  \e_{ab} Z \,,  \left[ P_a, B_b \right] =  -\e_{ab} Y \,,
\label{algebradouble}
\eea
where $a=1,2,$ and we used the three dimensional identities $J_{ab} = \e_{ab} J$ and $S_{ab} = \e_{ab} S$. This extended algebra can be equipped with an invariant non-degenerate bi-linear form
\bea
&&(P_a,B_b) =  (G_a,T_b) = \delta_{ab}\,, \nonumber\\
&& (J,Y) = -(M,S)  = -(H,Z) = 1\,,
\label{bilinear}
\eea
which we can use to construct a Chern-Simons action
\bea
S = \frac{k}{4\pi} \int \rm{Tr} \left(A \wedge dA + \frac23 A \wedge A \wedge A \right) \,,
\label{TrAction}
\eea
where the gauge field $A = A_\m dx^\m$ is given by
\bea
A_\m&=& \t_\m H + e_\m{}^a P_a + \o_\m J + \o_\m{}^a G_a + m_\m M + s_\m S \nonumber\\
&& + t_\m{}^a T_a + b_\m{}^a B_a + y_\m Y + z_\m Z \,.
\eea
In its component form, this non-relativistic action is given by
\bea
S_{\rm{ENG}} &=& \frac{k}{4\pi} \int  d^3 x \, \ve^{\m\n\r} \Big(  e_\m{}^a R_{\n\r a}(B) + t_\m{}^a R_{\n\r a} (G) \nonumber\\
&& + y_\m R_{\n\r}(J) - m_\m R_{\n\r}(S) - \t_\m R_{\n\r}(Z)\Big) \,,
\label{action}
\eea
where we have used the curvatures
\bea
R_{\m\n}{}^a (B)  & =& 2  \partial_{[\m} b_{\n]}{}^a  + 2  \e^{ab} \o_{[\m }\ b_{\n]b} +  2\e^{ab}s_{[\m}\ \o_{\n]b} \,,  \nonumber\\
R_{\m\n}{}^a (G)&=& 2  \partial_{[\m} \o_{\n]}{}^a  + 2 \e^{ab} \o_{[\m}\ \o_{\n]b}\,,    \nonumber\\
R_{\m\n}(J) &=& 2  \partial_{[\m} \o_{\n]}\,,  \nonumber\\
R_{\m\n}(S) & =& 2  \partial_{[\m} s_{\n]} + \e_{ab} \o_{[\m}{}^a\  \o_{\n]}{}^b\,,  \nonumber\\
R_{\m\n}(Z) &=& 2  \partial_{[\m} z_{\n]} + 2 \e^{ab} \o_{[\m a}\ b_{\n] b}\,.
\eea
We refer the action (\ref{action}) as the extended Newtonian gravity. Although the extended Newtonian gravity is based on the central extension of the algebra that gives rise to an action for Newtonian gravity, it is quite distinct from the Newtonian gravity action of \cite{NewtonAction}. This can most easily be seen by the identity that connects the Riemann tensor to the group theoretical curvatures $R_{\m\n}(J)$ and $R_{\m\n}{}^a (G)$
\bea
R^\s{}_{\r\m\n} = \e^{ab} e^\s{}_a  e_{\r b} R_{\m\n} (J) - \e^{ab} e^\s{}_a  \t_\r R_{\m\n b} (G) \,.
\eea
This identity implies that the time component of the Ricci tensor is given by
\bea
\t^\m \t^\n R^\r{}_{\m\r\n} = - \e^{ab} \t^\n  e^\m{}_a  R_{\m\n b} (G) \,,
\label{identity}
\eea
which is precisely the leading term of the action for the Newtonian and extended Bargmann gravity \cite{EBG,EMG0,EBG2} up to a redefinition $\o^a \rightarrow \e^{ab} \o_b$. In three dimensions, this structure requires the invariant bi-linear form to have a non-zero $(P_a, G_b)$ component. On the other hand, such a non-zero component is forbidden by the algebra (\ref{algebradouble}) as $S$ is not a central charge and most importantly it doesn't commute with $P_a$ and $G_a$. This implies that the non-relativistic models do not follow the relativistic analogy such that there is a unique two-derivative theory that admit Poincar\'e symmetries, namely the Einstein-Hilbert action, with both a metric and a Chern-Simons formulation. In non-relativistic setting, the symmetry algebra of the Newtonian gravity of \cite{NewtonAction} does not admit an invariant bilinear form without a central extension. Once the algebra is extended, the resulting Chern-Simons model is distinct from the Newtonian gravity action.


The extended Newtonian gravity is not the Chern-Simons formulation of the Newtonian gravity of \cite{NewtonAction} and yet it might be expected that it couples to matter in the same way that the Newtonian gravity couples to matter, i.e., its matter coupling gives rise to
\bea
\t^\m \t^\n R^\r{}_{\m\r\n} = - \e^{ab} \t^\n  e^\m{}_a  R_{\m\n b} (G) \propto j^0 \,,
\label{KeyRelation}
\eea
as an equation of motion which is the Poission equation for Newtonian gravity.  In what follows we show the matter coupling of the extended Newtonian algebra resembles to that of extended Bargmann gravity \cite{EBG} and admits backgrounds with non-trivial curvature whenever matter is present. When there is no matter coupling, the $z_\m$, $b_\m{}^a$ and $s_\m$ equations of motions for the extended Newtonian gravity (\ref{action}) imply the standard curvature constraints
\bea
R_{\m\n}(H) = 0\,, \quad R_{\m\n}{}^a (P) = 0 \,, \quad R_{\m\n}(N) = 0 \,,
\eea
where 
\bea
R_{\m\n} (H) &=& 2  \partial_{[\m} \t_{\n]} \,, \nonumber\\
R_{\m\n}{}^a (P) &=& 2  \partial_{[\m} e_{\n]}{}^a  - 2 \e^{ab} \o_{[\m b}\  \t_{\n]} + 2 \e^{ab} \o_{[\m}\ e_{\n]b}\,,  \nonumber\\
R_{\m\n}(N) &=& 2  \partial_{[\m} m_{\n]} + 2 \e^{ab} \o_{[\m a} e_{\n]b}\,.
\eea
Here, first equation implies that the spacetime can be foliated in an absolute time direction, while the last two equations identifies $\o_\m$ and $\o_\m{}^a$ in terms of $\t_\m, e_\m{}^a$ and $m_\m$. Hence the action (\ref{action}) provides a model that is defined on torsionless Newton-Cartan geometry. When the matter coupling is included, however, any matter couplings to that includes $z_\m$ would change the foliation constraint $R_{\m\n}(H) = 0$. For simplicity let us consider the matter Lagrangian $\cL_{m}$ to be independent of $z_\m$ and consider the following action
\bea
S &=& S_{\rm{ENG}} + \int d^3 x\, e\, \cL_m \nonumber\,,
\label{matteraction}
\eea
We can look at the variation of the action with respect to $t_\m{}^a$ as it determines how the matter coupling would appear on the right-hand side of the Poisson equation via the identity (\ref{identity}). The $t_\m{}^a$ field equation read
\bea
e^{-1} \varepsilon^{\m\n\r} R_{\m\n a}{} (G) =  - \frac{4\pi}{k}T^\rho{}_a \,,    
\eea
where
\bea
T^\r{}_a = e^{-1} \frac{\delta}{\delta t_\r{}^a } \left(e \cL_m \right) \,.
\eea
This indicates that the time component of the Ricci tensor is given by
\bea
\t^\m \t^\n R^\r{}_{\m\r\n} \propto e_\m{}^a T^\m{}_a \,.
\eea
The Newtonian gravity of \cite{NewtonAction} yields an equation for the Ricci tensor such that only the purely time-like component of the Ricci tensor is non-zero, i.e. $\t^\m \t^\n R^\r{}_{\m\r\n} \propto j^0$. For the extended Newtonian gravity, the  matter sources all components of the Riemann tensor similar to extended Bargmann gravity. Consequently, extended Newtonian gravity admits backgrounds with non-trivial curvature whenever matter is present.

Being a Chern-Simons theory, one of the immediate conclusions that can be made regarding the extended Newton gravity that it has no propagating degree of freedom. From a contraction viewpoint, this implies that the extended Newton gravity should be a non-relativistic limit of a non-interacting bi-metric gravity model due to its field content. This was noted in \cite{NewtonAction} to some extend that the $D$-dimensional analogue of the algebra (\ref{algebradouble}) without the central extensions can be obtained from a contraction of the direct sum of the Poincare and Euclidean algebras in $D$ dimensions. The non-interacting model that we consider here is thus the sum of the Einstein gravity in Lorentzian and the Euclidean signatures plus a Chern-Simons action for two abelian gauge fields  $Z_1$ and $Z_2$
\bea
S &=& \frac{k\l}{4 \pi } \int  \varepsilon^{\m\n\r} \Big( \eta_{AB}  E_{1 \mu}^A R_{\n\r}^B(\O_1)  +  \delta_{AB}  E_{2 \mu}^A R_{\n\r}^{B}(\O_2) \nonumber\\
&& \qquad \qquad \quad  + 2Z_{1 \m} \partial_\n Z_{2 \r} \Big) \,.
\label{EZDG}
\eea
where $\eta_{AB}$ is Minkowski metric ($\eta_{AB} = {\rm diag}(-1,1,1)$), $\delta_{AB}$ is Euclidean metric ($\delta_{AB} = {\rm diag}(1,1,1)$) and $E_{\m I}^A$ and $\O_{\m I}^A$ with $I=1,2$ represent the dreibein and spin connections respectively. Here, the curvatures are given by
\bea
R_{ \n\r}^A(\O_1) =  2  \partial_{[\m} \O_{1 \n]}^A  + \epsilon^{ABC} \O_{1 [\m B} {} \O_{1 \n]C} \,,\nonumber\\
R_{ \n\r}^A(\O_2) =  2  \partial_{[\m} \O_{2 \n]}^A  + \tilde{\epsilon}^{ABC} \O_{2 [\m B} {} \O_{2 \n]C} \,,
\eea
with the following convention for the Levi-Civita symmbols  
\bea
\varepsilon_{012} = -\varepsilon^{012} = -1\,, \qquad  \tilde{\varepsilon}_{012} = \tilde{\varepsilon}^{012} = 1 \,.
\eea
For the contraction procedure, we express the fields $E_{\m I}^A,\O_{\m I}^A, Z_{1\m}, Z_{2\m}$ as
\bea
 E_{1 \m}^a = \l^2 \sqrt{2}e_\m{}^a +\frac{\sqrt{2}}{2} t_\m{}^a  \,, & \quad  & E_{2\m}^a = \l^2 \sqrt{2} e_\m{}^a - \frac{\sqrt{2}}{2} t_\m{}^a \,,\nonumber\\
 \O_{1\m}^a = \frac{1}{2\sqrt{2}\l^3} b_\m{}^a + \frac1{\l} \o_\m{}^a \,,  &\quad  & \O_{2\m}^a = \frac{1}{2\sqrt{2}\l^3} b_\m{}^a - \frac1{\l} \o_\m{}^a \,,\nonumber\\
 E_{1\m}^0 = -\frac{1}{\l} y_\m + 2\l^3 \t_\m \,, & \quad  &E_{1\m}^0 = - 2\l m_\m  + 2\l^3 \t_\m \,,\nonumber\\
  \O_{1\m}^0 = \o_\m + \frac{1}{\l^2} s_\m \,, & \quad  &\O_{2\m}^0 = -\o_\m - \frac{1}{2\l^4} z_\m \,,\nonumber\\
   Z_{1\m} =- \frac{1}{\l}m_\m+ 2\l \t_\m  & \quad &Z_{2\m} = s_\m + 2 \l^2 \o_\m \,.
\eea
Using these expressions in the action (\ref{EZDG}) and taking the limit $\l \rightarrow \infty$ we precisely recover the extended Newtonian gravity (\ref{action}). 

As mentioned, another advantage of the Chern-Simons formulation is that it allows us to find the supersymmetric description of the extended Newtonian gravity. To find the supersymmetric extension of the algebra (\ref{algebradouble}) we proceed along the lines of \cite{EBG} and consider five fermionic generators $Q^{\pm}, W^{\pm}$ and $R$ that are all Majorana spinors. With the assistance of the computer algebra program \textit{Cadabra} \cite{Cadabra1,Cadabra2}, we find that the superalgebra has the following additional non-zero commutators
\bea
\{ Q^+_\a ,  R_\b  \} = (\g_0 C^{-1})_{\a \b} M ,\, \quad && [J, Q^\pm ] = - \frac{1}{2} \g_0 Q^\pm  \,,\nonumber\\
\{ Q^+_\a ,  Q^+_\b  \} = (\g_0 C^{-1})_{\a \b} H  ,\, \quad && [J, R ] = - \frac{1}{2} \g_0 R    \,,\nonumber\\
  \{ Q^+_\a ,  Q^-_\b  \}  = -(\g_a C^{-1})_{\a \b} P^{a}   ,\, \quad   && [J, W^\pm ] = - \frac{1}{2}  \g_0 W^\pm \,,\nonumber\\
  \{ Q^+_\a ,  W^+_\b  \}  = (\g_0 C^{-1})_{\a \b} Y\,, \quad &&  [G_a, Q^+] = - \frac{1}{2} \g_a Q^- \,,\nonumber\\
   \{ Q^+_\a ,  W^-_\b  \} = (\g_a C^{-1})_{\a \b} T^a    ,\, \quad && [G_a, Q^-] = - \frac{1}{2} \g_a R  \,,\nonumber\\
 \{ Q^-_\a ,  Q^-_\b  \} = (\g_0 C^{-1})_{\a \b} M\,, \quad    && [G_a, W^-] = - \frac{1}{2} \g_a W^+ \,,\nonumber\\
    \{ Q^-_\a ,  R_\b  \} = - (\g_a C^{-1})_{\a \b} T^a ,\, \quad && [G_a, R] =  \frac12 \g_a W^- \,,\nonumber\\
   \{ Q^-_\a ,  W^-_\b  \} = (\g_0 C^{-1})_{\a \b} Y \,, \quad    && [S, Q^+] = -\frac{1}{2}\g_0 R \,,\nonumber\\
    \{ R_\a ,  R_\b  \} = -(\g_0 C^{-1})_{\a \b} Y  \,, \quad && [S, Q^-] =  \frac12 \g_0 W^-  \,,\nonumber\\
    \left[Z, Q^+ \right] =  \frac12 \g_0 W^+   \,, \quad  && [B_a, Q^\pm] = \frac12 \g_a W^\mp   \,, \nonumber\\
    \left[S, R\right] =\frac12   \g_0 W^+ \,. \quad &&
\label{Superalgebradouble}
\eea
Note that $Z$ is no longer a central charge in the supersymmetric extension of the algebra. The invariant supertrace is given by (\ref{bilinear}) extended with
\bea
&& (Q^+_\a, W^+_\b) = -2 (C^{-1})_{\a \b}\,, \quad  (Q^-_\a, W^-_\b) = -2 (C^{-1})_{\a \b}\,, \nonumber\\
&& (R_\a, R_\b) =  2 (C^{-1})_{\a \b} \,.\label{Superbilinear}
\eea
Introducing the gauge field
\bea
A_\m &=& \t_\m H + e_\m{}^a P_a + \o_\m J + \o_\m{}^a G_a + m_\m M + s_\m S \nonumber\\
&& + t_\m{}^a T_a + b_\m{}^a B_a + y_\m Y + z_\m Z + \bar\psi^+_\m Q^+ \nonumber\\
&&  +  \bar\psi^-_\m Q^- + \bar\r_\m R + \bar\phi^+_\m W^+ +  \bar\phi^-_\m W^-
\eea
the action for the extended Newtonian supergravity read
\bea
S &=& \frac{k}{4\pi} \int  d^3 x \, \ve^{\m\n\r} \Big(  e_\m{}^a R_{\n\r a}(B) + t_\m{}^a R_{\n\r a} (G) \nonumber\\
&&   + y_\m R_{\n\r}(J)  - m_\m R_{\n\r}(S)  - \t_\m R_{\n\r}(Z) \nonumber\\
&&- \bar\psi_\m^+ R_{\n\r}(W^+) - \bar\phi_\m^+ R_{\n\r}(Q^+) -  \bar\psi_\m^- R_{\n\r}(W^-) \nonumber\\
&&    -  \bar\phi_\m^- R_{\n\r}(Q^-) + \bar \r_\m R_{\n\r} (R)\Big) \,.
\label{superaction}
\eea
Here the supercovariant curvatures are given by
\bea
 R_{\m\n}(Q^+) &=& 2 \partial_{[\m} \psi_{\n]}^+ +  \o_{[\m} \g_0 \psi_{\n]}^+  \,,\nonumber\\ 
 R_{\m\n}(Q^-) &=& 2 \partial_{[\m} \psi_{\n]}^- +  \o_{[\m} \g_0 \psi_{\n]}^-  + \o_{[\m}{}^a \g_a \psi_{\n]}^+  \,,\nonumber\\ 
 R_{\m\n}(R) &=& 2 \partial_{[\m} \rho_{\n]} +  \o_{[\m} \g_0 \rho_{\n]}  + \o_{[\m}{}^a \g_a \psi_{\n]}^- +  s_{[\m} \g_0 \psi_{\n]}^+ \,, \nonumber\\ 
 R_{\m\n}(W^+) &=& 2 \partial_{[\m} \phi_{\n]}^+ +  \o_{[\m} \g_0 \phi_{\n]}^+  + \o_{[\m}{}^a \g_a \phi_{\n]}^- -  s_{[\m} \g_0 \r_{\n]} \nonumber\\
&&-  z_{[\m} \g_0 \psi_{\n]} ^+  - b_{[\m}{}^a \g_a \psi_{\n]} ^-\,,\nonumber\\ 
R_{\m\n}(W^-) &=& 2 \partial_{[\m} \phi_{\n]}^- +  \o_{[\m} \g_0 \phi_{\n]}^-  - \o_{[\m}{}^a \g_a \r_{\n]} \nonumber\\
&& -  s_{[\m} \g_0 \psi_{\n]}^- -   b_{[\m}{}^a \g_a \psi_{\n]} ^+\,. \label{supercurvatures}
\eea
These curvatures transform covariantly with respect to the supersymmetry transformation rules
\bea
\d \t_\m &=& - \bar{\e}^+ \g_0 \psi^+_\mu \,,\nonumber\\ 
\d e_\m{}^a &=&  \bar{\e}^+ \g^{a} \psi_\m^- +  \bar{\e}^- \g^{a} \psi_\m^+ \,,\nonumber\\ 
\d m_\m &=& - \bar{\e}^- \g_0 \psi_\m^- -  \bar{\e}^+ \g_0 \rho_\m -  \bar{\eta} \g_0 \psi_\m^+ \,,\nonumber\\
\d y_\m &=& - \bar{\e}^+ \g_0 \phi_\m^+ -  \bar{\chi}^+ \g_0 \psi_\m^+  -  \bar{\e}^- \g_0 \phi_\m^- -  \bar{\chi}^- \g_0 \psi_\m^- \nonumber\\
&& +  \bar{\eta} \g_0 \r_\m \,,\nonumber \\
\d t_\m ^a&=& - \bar{\e}^+ \g^a \phi_\m^- - \bar{\chi}^- \g^a \psi_\m^+ +  \bar{\e}^- \g^a \r_\m +  \bar{\eta} \g^a \psi_\m^- \,,\nonumber \\
\d \psi_\m^+ &=& \partial_\m \e^+ + \frac12 \o_\m \g_0 \e^+ \,, \nonumber \\
\d \psi_\m^- &=&  \partial_\m \e^- + \frac12 \o_\m \g_0 \e^- + \frac12 \o_\m{}^a \g_a \e^+ \,, \nonumber \\
\d \r_\m &=&  \partial_\m \eta + \frac12 \o_\m \g_0 \eta + \frac12 \o_\m{}^a \g_a \e^- + \frac12 s_\m \g_0 \e^+  \,, \nonumber \\
\d \phi_\m^+ &=&  \partial_\m \chi^+ + \frac12 \o_\m \g_0 \chi^+ + \frac12 \o_\m{}^a \g_a \chi^- - \frac12s_\m \g_0 \eta \nonumber\\
&& -  \frac12z_\m \g_0 \e^+ - \frac12b_\m{}^a  \g_a \e^-  \,,\nonumber\\
\d \phi_\m^- &=&  \partial_\m \chi^- + \frac12 \o_\m \g_0 \chi^- -   \frac12\o_\m{}^a \g_a \eta -  \frac12s_\m \g_0 \e^- \nonumber\\
&&-  \frac12b_\m{}^a  \g_a \e^+   \,,
\eea
where $\e^\pm, \chi^\pm$ and $\eta$ are the parameters of the local $Q^\pm, W^\pm$ and $R$ transformations, respectively.

In this paper, we presented an action principle for a particular three-dimensional non-relativistic gravity that is invariant under the central extension of the symmetry group that leaves the Newtonian gravity action invariant. The model is distinct from the Newtonian gravity both at the level of action and the matter coupling. By choosing fields appropriately, we show that this action can be obtained by a contraction procedure. Our model is of the Chern-Simons type which allowed us to establish the supersymmetric completion by extending the algebra with five supersymmetry generators.

The algebra that we present here can be extended to accommodate a cosmological constant. In this case, we expect that there are other invariant bi-linear forms, in particular, there might be an ``exotic" extended Newtonian gravity that arises from the contraction of the sum of the ``exotic" Einstein gravity in Lorentzian and the Euclidean signatures \cite{3d1}. Furthermore, the algebra can also be extended to accommodate scale and Schr\"odinger symmetry along the lines of \cite{EBG2}. Once these extensions are established, it would also be interesting to see whether there is a further extension to include supersymmetry. 

The extended Newtonian gravity has no propagating degree of freedom as it is a Chern-Simons gauge theory. Degree of freedom can be introduced by considering higher-derivative extensions or bi-gravity models. This can be most easily achieved by promoting our Chern-Simons theory to a Chern-Simons-like theory, i.e., the action is still of the form (\ref{TrAction}) but $A_\m^I$ is no longer has to be a gauge field \cite{CSL1}. While higher derivative models are expected to modify the Poisson equation for Newtonian gravity, the bi-gravity models are particularly relevant to a geometric interpretation of GMP states by providing a particular spin-2 planar Schr\"odinger equation \cite{GMP,GMP2,GMP3,GMP4,GMP5,GMP6}. Thus, it would very be interesting to see if one can construct non-relativistic interacting bi-gravity theories with the correct properties in light of the discussions that we provide here. Being a non-linear theory, it would also be interesting to develop a Hamiltonian formulation to analyze the degree of freedom of the resulting higher derivative or bi-gravity models in Chern-Simons like formalism.

Finally, the superalgebra that we presented in this paper is constructed by hand but obtained not from a contraction. In principle, this could be possible by considering the contraction of the sum of Poincar\'e and Euclidean supergravity along with an action for an abelian vector multiplet. The analysis of the matter multiplets, which can either be obtained by direct construction or by contraction, would be very interesting as they have important applications in supersymmetric non-relativistic field theories by means of localization.

\section{Acknowledgments}

We are greatful to Jan Rosseel for comments and suggestions. NO and UZ are supported in parts by Istanbul Technical University Research Fund under grant number TDK-2018-41133. The work of MO is supported in part by TUBITAK grant 118F091.

\providecommand{\href}[2]{#2}\begingroup\raggedright\endgroup

\end{document}